\documentclass[twocolumn,floats,floatfix,prd,superscriptaddress,nofootinbib,longbibliography]{revtex4-1}
\usepackage{graphicx,epsfig}
\usepackage{amssymb,amsmath,amsthm,amsfonts}
\usepackage{bm}
\usepackage[inline]{enumitem}
\usepackage{tensor}
\usepackage[linktocpage]{hyperref}
\usepackage[caption=false]{subfig}
\usepackage[usenames,dvipsnames]{xcolor}
\usepackage{url}
\usepackage[inline]{enumitem}
\usepackage{xspace}
\usepackage{comment}
\usepackage{cancel}
\usepackage{multirow}

\def\B{\mathcal{B}}

\def\bdot{\boldsymbol{\cdot}}

\def\tUps{\tilde{\Upsilon}}

\def\Rp{\mathcal{R}^+}

\def\B{\mathcal{B}}

\def\hPsi{\hat{\Psi}}
\def\hK{\hat{K}}
\def\hH{\hat{H}}

\def\bfR{\mathbf{R}}
\def\bfF{\mathbf{F}}

\def\I{i\,}
\def\k5{\kappa_5}

\def\newacronym#1#2#3{\gdef#1{\gdef#1{#2\xspace}#3 (#2)\xspace}}
\newacronym{\bh}{BH}{black hole}
\newacronym{\mbh}{MBH}{magnetized black hole}
\newacronym{\ts}{TS}{topological star}
\newacronym{\bbh}{BBH}{black hole binary}
\newacronym{\gr}{GR}{General Relativity}
\newacronym{\bssn}{BSSN}{Baumgarte-Shapiro-Shibata-Nakamura}
\newacronym{\ts}{TS}{topological star}
\newacronym{\qnm}{QNM}{quasinormal mode}
\newacronym{\emd}{EMD}{Einstein-Maxwell-dilaton}
\newacronym{\nr}{NR}{Numerical Relativity}

\newcommand{\sapienza}{Dipartimento di Fisica, Sapienza Università 
	di Roma, Piazzale Aldo Moro 5, 00185, Roma, Italy}
\newcommand{\infn}{INFN, Sezione di Roma, Piazzale Aldo Moro 2, 00185, Roma, Italy}

\begin{document}

\title{Nonradial stability of topological stars}

\author{Alexandru Dima}
\email{alexandru.dima@uniroma1.it}
\affiliation{\sapienza}
\affiliation{\infn}

\author{Marco Melis}
\email{marco.melis@uniroma1.it}
\affiliation{\sapienza}
\affiliation{\infn}

\author{Paolo Pani}
\email{paolo.pani@uniroma1.it}
\affiliation{\sapienza}
\affiliation{\infn}

\begin{abstract}
    Topological stars are regular, horizonless solitons arising from dimensional compactification of Einstein-Maxwell theory in five dimensions, which could describe qualitative properties of microstate geometries for astrophysical black holes. They also provide a compelling realization of ultracompact objects arising from a well-defined theory and display all the  phenomenological features typically associated with black hole mimickers, including a (stable) photon sphere, long-lived quasinormal modes, and echoes in the ringdown.
    By completing a thorough linear stability analysis, we provide strong numerical evidence that these solutions are stable against nonradial perturbations with zero Kaluza-Klein momentum. 
\end{abstract}

\maketitle

\tableofcontents

\begin{figure*}[th]
  \centering
\includegraphics[width=0.495\textwidth]{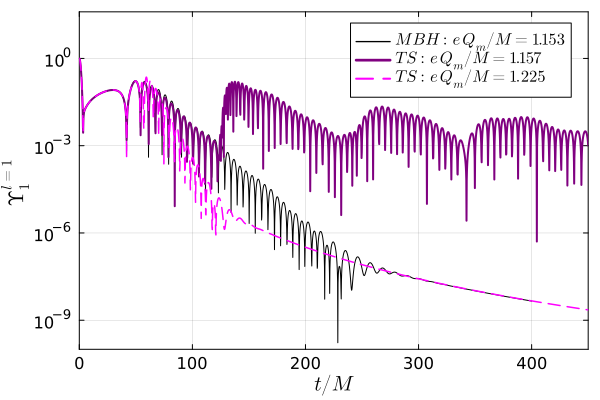}
\includegraphics[width=0.495\textwidth]{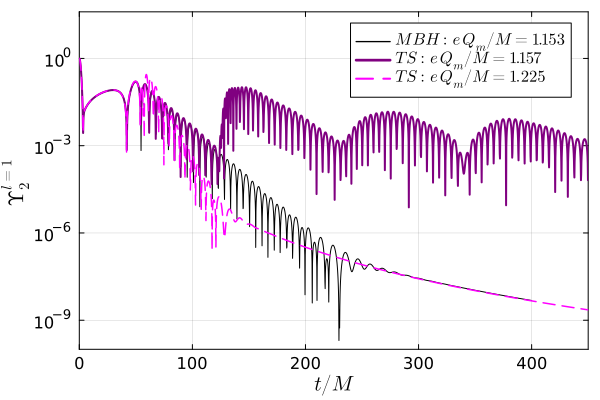}
  \caption{Comparison between the linear response of a nearly-extremal magnetized BH (black) and a second-kind TS with same mass and similar charge-to-mass ratio (purple) to $l=1$ Type-II perturbations (odd-parity electromagnetic, and even-parity scalar). The left and right panels refer to scalar-driven and electromagnetic-driven perturbations, respectively. 
  For comparison we show also the case of a first-kind TS (magenta), which displays a different prompt ringdown and no long-lived modes.
  } \label{fig:Ups12_RingdownVSEchoes}
\end{figure*}

\section{Introduction}
An outstanding challenge related to black holes~(BHs) is providing a microscopic interpretation of their entropy, which is famously equal to the horizon area in Planck units~\cite{Bekenstein:1973ur}, and consequently enormous for astrophysical BHs.
A compelling explanation for this entropy is given by the fuzzball paradigm of string theory, according to which a classical BH can be interpreted as a thermodynamical approximation of a huge number of regular quantum states~\cite{Mathur:2009hf,Bena:2022rna,Bena:2022ldq}. 
In the classical limit, the latter are described by ``microstate geometries'': solitons with the same mass and charges as a BH, but with a drastically different inner geometry, wherein the horizon is replaced by a smooth horizonless cap~\cite{Bena:2006kb,Bena:2016ypk,Bena:2017xbt,Bah:2021owp,Bah:2022yji}.
The absence of an event horizon in the fundamental description automatically resolves Hawking's information loss paradox~\cite{Hawking:1976ra,Polchinski:2016hrw}, since information can be retrieved by the horizon structure provided by the microstate geometries.

Two key ingredients of the fuzzball construction are the intrinsically higher dimensionality of the microstate solutions and their nontrivial topologies, which prevent the horizon-scale structure from collapsing.
While the microstates characterizing horizon-scale structure are smooth, topologically nontrivial geometries in ten dimensions, in four dimensions
they appear to have curvature singularities~\cite{Balasubramanian:2006gi,Bena:2022fzf}.
These solutions are particularly involved and, from the four-dimensional perspective, do not have any spatial isometries. This has so far prevented dynamical studies, which would instead be highly informative about the linear stability and out-of-equilibrium dynamics of these solutions (see Ref.~\cite{Ikeda:2021uvc} for an exception considering a test scalar field).

A particularly simple model incorporating some of the key fuzzball ideas was proposed some years ago in~\cite{Bah:2020ogh}, which discovered regular solitons in
Einstein-Maxwell theory in five dimensions. These solutions were dubbed topological stars~(TSs) since they have topological cycles supported by the magnetic flux.
TSs have several appealing features. Their four-dimensional reduction (upon compactifying the fifth dimension on a circle) asymptotically resembles magnetic BHs, but lacks an event horizon, while the 5D counterpart remains free of curvature singularities. 
They can be macroscopically large compared to the size of the Kaluza-Klein circle so they could describe qualitative properties of microstate geometries for astrophysical BHs.
Furthermore, their gravitational potential smoothly interpolates between that of a moderately compact star and that of a ultracompact object which could mimic the peculiar phenomenological features of BHs~\cite{Cardoso:2019rvt}.
Finally, extensions of these solutions can be constructed such that they are globally neutral and asymptote to a Schwarzschild BH~\cite{Bah:2022yji}, although at the expenses of introducing more complexity and breaking spherical symmetry.

The spherical symmetry of the simplest TSs is also particularly convenient to study the properties of these solutions, for example performing ray tracing~\cite{Heidmann:2022ehn}, studying the dynamics of test fields~\cite{Heidmann:2023ojf,Bianchi:2023sfs} and, recently, their linear response~\cite{Dima:2024cok,Bena:2024hoh,Bianchi:2024vmi,Bianchi:2024rod}.
In particular, Refs.~\cite{Dima:2024cok,Bena:2024hoh} studied the linear dynamics of TSs, to compute their quasinormal modes~(QNMs) and ringdown waveform.
The presence of a background
magnetic field
mixes perturbations with opposite parity, but still allows for separating them in two independent sectors, dubbed Type~I and Type~II (see below for details).
The Type~I sector was found to be free of instabilities.
The Type~II sector is significantly more involved and only radial perturbations (that belong to this sector) were analyzed in~\cite{Dima:2024cok,Bena:2024hoh}.
In this case there exists a radial unstable mode with momentum along the fifth direction, which is directly connected to the Gregory-Laflamme instability~\cite{Gregory:1993vy} of charged black strings~\cite{Miyamoto:2006nd,Stotyn:2011tv,Bah:2021irr}. However, this mode only affects TSs with small compactness, leaving the most interesting region of the parameter space available.

Here we continue the linear stability of TSs by analyzing the nonradial Type~II perturbations with zero Kaluza-Klein momentum. As we shall discuss, also in this sector we do not find any signature for instability, suggesting that TSs are stable against nonradial perturbations.

A representative example of our analysis is shown in Fig.~\ref{fig:Ups12_RingdownVSEchoes}, which compares the time-domain linear response of a magnetized BH with different families of TSs. As we shall discuss in detail, the ringdown of very compact (second-kind, see classification below) TSs is initially very similar to that of a BH and displays slowly-decaying echoes~\cite{Cardoso:2016oxy,Cardoso:2016rao,Cardoso:2017cqb,Dima:2024cok} at late time, whereas less compact (first-kind) TSs display a standard exponentially suppressed ringdown, but with different modes compared to the BH case.

\section{Theoretical framework}
We consider Einstein-Maxwell theory in 5D,
\begin{equation}
    S_5 = \int d^5 x \sqrt{-\mathbf{g}} \left(\frac{1}{2\kappa_5^2}\bfR - \frac{1}{4}\bfF_{AB}\bfF^{AB} \right)\,.
\end{equation}
Differently from what was done in~\cite{Dima:2024cok}, here we do not perform a four-dimensional compactification, and keep working in the 5D picture. Thus, we consider the TS and magnetized BH found in~\cite{Bah:2020ogh} as our background solutions, which are described by the five-dimensional line element\footnote{Although we work in 5D, we will always have in mind a Kaluza-Klein compactification of the $y$ dimension on a circle with radius $R_y$. Having in mind their 4D version, we will refer to these solutions as (magnetized) BHs and TSs, although in 5D they are actually strings.}
\begin{align}\label{eq:BH/TS_5D_metric}
ds^2 & =-f_S dt^2+f_Bdy^2+\frac{1}{h}dr^2+r^2d\Omega_2^2
\\
F & = P \sin\theta \,d\theta \wedge d\phi
\end{align}
where
\begin{align}
&f_S=1-\frac{r_S}{r}\,, \hspace{1em}f_B=1-\frac{r_B}{r}\,,
\nonumber\\
& h=f_B f_S\,,\hspace{1em} P = \pm \frac{1}{\kappa_5}\sqrt{\frac{3r_Sr_B}{2}}\,.
\end{align}

The mass and magnetic charge of the solution are $M = \frac{2\pi}{\kappa_4^2}(2r_S+r_B)$ and $Q_m = \frac{1}{\kappa_4}\sqrt{\frac{3}{2}r_Sr_B}$, respectively, where we introduced the effective coupling constant in 4D defined as $\kappa_4^2:=\kappa_5^2/(2\pi R_y)$, with $R_y$ the radius of the compact extra dimension\footnote{See~\cite{Bah:2020pdz} for a discussion about the relative size of $r_B$ and $R_y$.}. 
The TS solution corresponds to the region $r_B>r_S$, whereas $r_B<r_S$ corresponds to a magnetized BH with event horizon located at $r=r_S$. For a TS, the metric compactified to 4D diverges at $r=r_B$, which corresponds to a curvature singularity, but the solution is perfectly regular in the five dimensional uplift.

We can classify TSs in two families, according to the number of photon spheres in their spacetime~\cite{Heidmann:2023ojf,Bianchi:2023sfs,Lim:2021ejg}: if $3/2 < {r_B}/{r_S} \leq 2$, the solution has a single unstable photon sphere at the boundary $r_{\rm ph}^{(1)} = r_B$ (\emph{first-kind} TS), whereas if $1 \leq {r_B}/{r_S} \leq 3/2$ the solution (\emph{second-kind} TS) has a stable photon sphere at $r_{\rm ph}^{(2)} = r_B$ and an unstable one at $r_{\rm ph}^{(1)} = \frac32 r_S$, just like the magnetized BH solution.
Thus, second-kind TSs classify as ultracompact objects~\cite{Cardoso:2019rvt}: their stable photon-sphere can support long-lived QNMs~\cite{Cardoso:2014sna,Heidmann:2023ojf,Bianchi:2023sfs,Dima:2024cok,Bena:2024hoh}, and their response in the time domain is initially very similar to that of a BH with comparable charge-to-mass ratio, while their late time response is governed by echoes~\cite{Cardoso:2016oxy,Cardoso:2016rao,Cardoso:2017cqb,Heidmann:2023ojf,Dima:2024cok}, analogously to what observed for test scalar perturbations of microstate geometries~\cite{Ikeda:2021uvc}.

The linear perturbations of TSs have been derived in~\cite{Dima:2024cok,Bena:2024hoh}, to which we refer for all details (see also~\cite{Guo:2022rms,Guo:2023vmc}).
The so-called Type-I sector contains odd-parity gravitational perturbations coupled to even-parity electromagnetic perturbations, whereas the Type-II sector contains even-parity gravitational and scalar perturbations coupled to odd-parity electromagnetic ones\footnote{Alternatively, one could use an electromagnetic duality transformation to avoid coupling perturbations with opposite parity~\cite{Bena:2024hoh, Pereniguez:2023wxf}.}.
Here we focus on Type-II nonradial perturbations, that are more involved and never studied so far.

Due to the spherical symmetry of the background, it is convenient to expand the perturbations in a basis of (scalar, vector, tensor) spherical harmonics of even and odd parity~\cite{Dima:2024cok} (see Appendix~\ref{app:RW}).
Beside the spherical mode ($l=0$, where $l$ is the harmonic index) studied in~\cite{Dima:2024cok,Bena:2024hoh}, the next-to-simplest Type-II mode is the dipole ($l=1$). 
In this case, we introduce auxiliary dynamical fields $\lbrace\Upsilon_1\,,\Upsilon_2\rbrace$, which correspond, respectively, to scalar-driven and electromagnetic-driven perturbations; and, in addition, a constrained perturbation $\Lambda$ to close the system. One can derive (see Appendix~\ref{app:deteqs} for the derivation) the evolution equations for the two dynamical degrees of freedom corresponding to Type-II perturbations with $l=1$:
%
\begin{align}
  \label{eq:TypeII_l1}
  & {\cal D}[ \Upsilon_i]
  + \sum_{j=1,2} W_{ij} \partial_r \Upsilon_j
  - \sum_{j=1,2} V_{ij} \Upsilon_j
  = 0
\end{align}
where $i\,,j=1,2$ and we introduced the differential operator ${\cal D} =\omega^2 +\left(f_Bf_S^2\right) \partial_r^2$. The coefficients appearing in Eq.~\eqref{eq:TypeII_l1} are reported in extended form in Appendix~\ref{app:deteqs}.

For $l\geq2$, there are three dynamical degrees of freedom: a scalar, an odd-parity electromagnetic, and an even-parity gravitational mode. They are described by the equations:
\begin{align}
  \label{eq:TypeII_l2}
  & {\cal D}[ \tUps_i]
  + \sum_{j=1,2,3} \tilde{W}_{ij} \partial_r \tUps_j
  - \sum_{j=1,2,3} \tilde{V}_{ij} \tUps_j
  \nonumber\\
  &   + \sum_{j=1,2,3} \tilde{U}_{ij} \partial_r^2 \tUps_j
  = 0
\end{align}
where $\lbrace\tUps\rbrace_{i=1,2,3} =\lbrace \Phi, \B, \Rp \rbrace$ are the scalar-driven, electromagnetic-driven and gravity-driven perturbation fields. Once again, we refer to Appendix~\ref{app:deteqs} for the details about the definition of the variables, $\tUps_i$, the explicit expression of the coefficients, $\lbrace \tilde{U}_{ij}\,, \tilde{W}_{ij}\,, \tilde{V}_{ij}\rbrace$, and the derivation of Eqs.~\eqref{eq:TypeII_l2}.

\section{Numerical methods}

We solve the sets of equations~\eqref{eq:TypeII_l1} and~\eqref{eq:TypeII_l2} respectively for $l=1$ and $l=2$ Type-II perturbations both in the frequency domain, as an eigenvalue problem to compute the QNMs, and in the time domain, as an initial value problem to study the linear response of the solution.

We use the same boundary conditions and numerical techniques discussed in~\cite{Dima:2024cok}, to which we refer for more details.
In particular, both sets of equations~\eqref{eq:TypeII_l1} and~\eqref{eq:TypeII_l2} are Heun equations which possess a regular singular point at $r=r_B$. A Fuchsian analysis of the solutions expanded around the singular point allows us to identify the appropriate boundary conditions that are needed to ensure that linear perturbations on a TS background are regular~\cite{Dima:2024cok}.

In contrast to~\cite{Dima:2024cok}, where the perturbations equations were decoupled and of the Schrödinger-like form, in this case Eqs.~\eqref{eq:TypeII_l1} and~\eqref{eq:TypeII_l2} are coupled, both for $l=1$ and $l\geq2$. In this regard, for the computation of the QNMs we use a matrix-valued direct integration method as discussed in detail in~\cite{Pani:2013pma,Pani:2012bp}. For an array of perturbations $\lbrace\Upsilon\rbrace_{j=1,2}$ (similarly for $\lbrace\tUps\rbrace_{j=1,2,3}$) we consider the following series expansions near $r_B$ (the same holds for the magnetized BH case, near $r_S$) 
\begin{equation}
    \Upsilon^{(j)} = (r-r_B)^\lambda \sum_{i=0}^\infty c_i^{(j)} (r-r_B)^i\,, 
\end{equation}
where $\lambda$ is fixed by the Fuchsian analysis,
and we choose an orthogonal basis of unit vectors for the $N-$dimensional space, $c_0^{(1)}=(1,0,0,..,0)$, $c_0^{(2)}=(0,1,0,...,0)$, ..., $c_0^{(N)}=(0,0,0,...,1)$. At infinity the solution is given by a linear combination of ingoing and outgoing waves
\begin{align}
    \Upsilon^{(j)} \sim B^{(j)}\, e^{- i \omega r}\, r^{\Tilde{\lambda}} +  C^{(j)}\, e^{i \omega r}\, r^{-\Tilde{\lambda}} \,,
\end{align}
where again $\Tilde{\lambda}$ is fixed by the Fuchsian analysis. To compute the QNMs we request only outgoing waves to survive. For the present case of a system of $N$ coupled second order ODEs we use the orthogonal basis of the $N-$dimensional space and after performing $N$ integrations from the inner boundary to infinity we construct the following $N \times N$ matrix
\begin{equation}
    \textbf{S}(\omega) = \begin{pmatrix}
        B_1^{(1)} & B_1^{(2)} & ... & B_1^{(N)} \\ 
        B_2^{(1)} & B_2^{(2)} & ... & ... \\
        ... & ... & ... & ... \\
        ... & ... & ... & ... \\
        B_N^{(1)} & ... & ... & B_N^{(N)}
    \end{pmatrix}
\end{equation}
and we impose $\det \textbf{S}(\omega) = 0$, which is the condition defining the QNMs.

\begin{figure}[th]
  \centering
\includegraphics[width=0.485\textwidth]{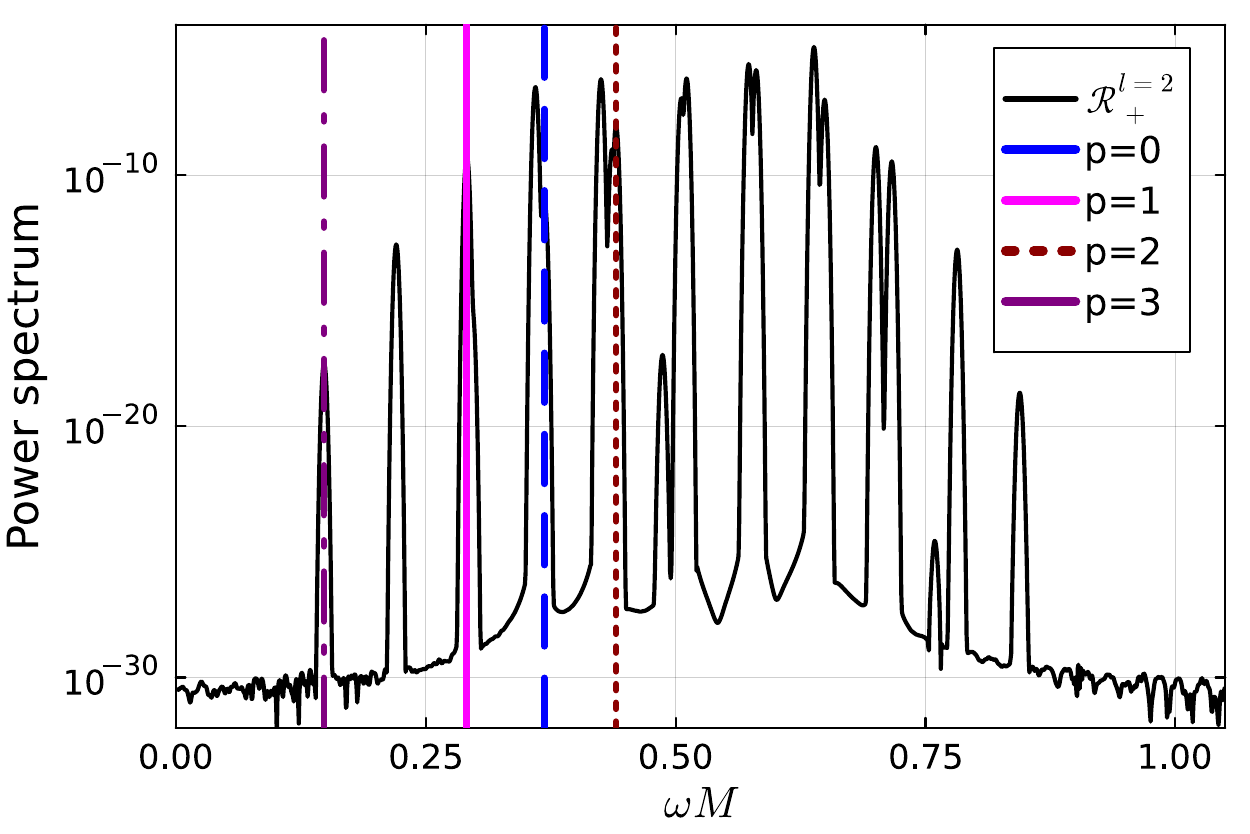}
  \caption{Power spectrum of Type-II perturbations with $l=2$ around a second-kind TS with $r_S/r_B=0.99$, showing the peaks corresponding to the long-lived modes reported in Table~\ref{tab:QNMs_TSs}, which are labeled by an integer $p=0,1,2,3$. The remaining modes are stable and sub-dominant (i.e. more damped) with respect to the labeled ones.} 
  \label{fig:PWS}
\end{figure}

For the time-domain computations, we follow the same approach as~\cite{Dima:2024cok}.  
The evolution equations are integrated using the method of lines. Spatial derivatives are discretized with a fourth-order finite-difference stencil, and time stepping is performed using a fourth-order Runge-Kutta method. At the outer boundary of the computational grid we impose outgoing radiative boundary conditions with a fourth-order finite difference approximation, which help reduce the numerical noise from the boundary. At the inner boundary, we carefully implemented boundary conditions to ensure regularity or radiative behavior depending on the nature of the object (namely TS or BH). We also incorporate techniques to improve numerical stability and accuracy, such as adaptive stretching of the computational domain and long simulations to capture long-lived modes~\cite{Dima:2024cok}. After obtaining the time-domain signal, QNMs are extracted by performing a Fourier transform and fitting the peaks in the frequency spectrum (see Fig.~\ref{fig:PWS} for an example). This approach ensures consistency between frequency-domain and time-domain computations, enabling detailed analysis of the linear response and stability of the studied systems. 

\section{Results}
In this section we present the linear-perturbation analysis of magnetized BHs and TSs, both in the frequency- and in the time-domain.
We order the QNMs accordingly to their imaginary part and label them with an integer $p=0,1,2,..$, with $p=0$ denoting the mode with the smallest imaginary part (in absolute value). Note that this classification does not differentiate among modes belonging to different families (i.e., gravitational-, electromagnetic-, or scalar-driven).

\begin{figure*}[th]
  \centering
  \includegraphics[width=1\textwidth]{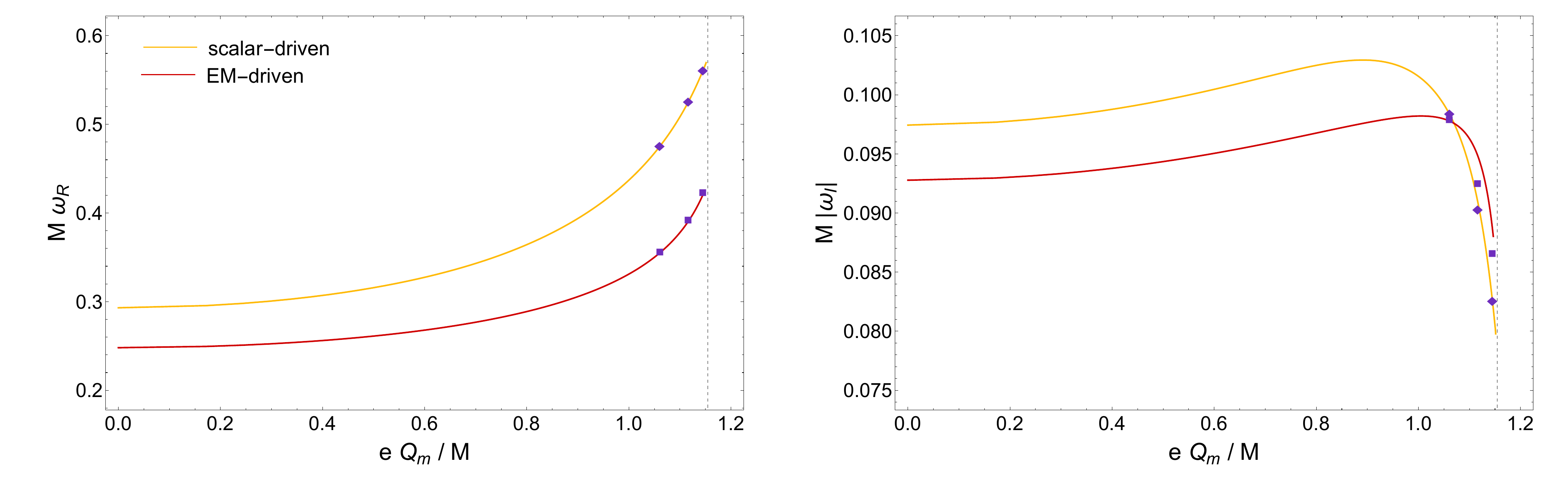}
  \includegraphics[width=1\textwidth]{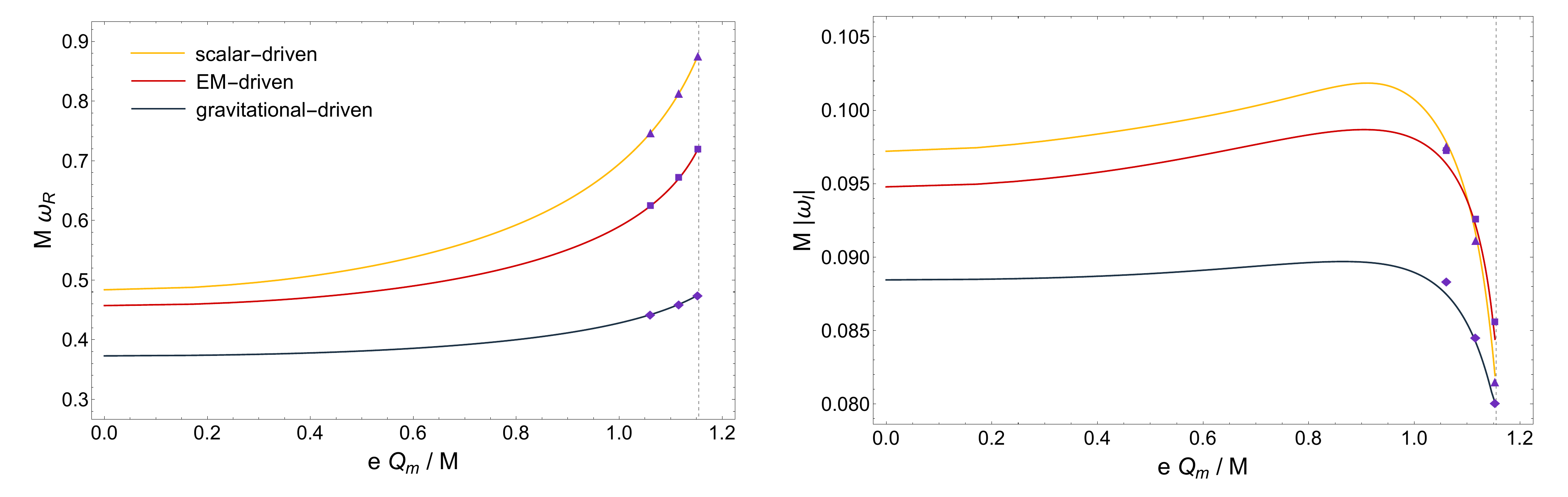}
  \caption{QNMs of a magnetized BH as a function of the magnetic charge for Type-II perturbations. Top panels refer to $l=1$, electromagnetic and scalar perturbations, while bottom panels refer to $l=2$ gravitational, electromagnetic, and scalar perturbations. The left (right) panels show the real (imaginary) part of the mode. In the limit $r_B \rightarrow 0$, i.e. $Q_m \rightarrow 0$, the frequencies are those of the fundamental QNMs for test scalar, vector and spin-2 tensor fields in the background of the Schwarzschild BH. The vertical dashed line denotes the extremal BH.
  The solid lines where obtained with the f-domain method, while
  the markers corresponds to the modes extracted from the Fourier transform of the time-domain signal. We classify the modes based on the hierarchy also in terms of $p$: for $l=1$, $p=0,1$ correspond to electromagnetic- and scalar-driven modes, respectively; whereas for $l=2$, $p=0,1,2$ correspond to gravitational-, electromagnetic- and scalar-driven modes, respectively. }\label{fig:QNMs_TypeII_l1l2_BH}
\end{figure*}

The results for magnetized BHs are summarized in Fig.~\ref{fig:QNMs_TypeII_l1l2_BH} for $l=1$ (top panels) and $l=2$ (bottom panels). For $l=2$, we have 3 different families of modes associated to the gravitational-, electromagnetic-, and scalar-driven perturbations in the neutral limit ($Q_m\to0$) which are all coupled to each other for nonvanishing $Q_m$. For $l=1$, we have the same effect both for electromagnetic- and scalar-driven modes only, since the gravitational ones do not propagate.
The markers shown in the plots corresponds to the modes extracted from the Fourier transform of the time-domain signal (see Fig.~\ref{fig:PWS} for an example of the power spectrum in the TS case, and Ref.~\cite{Dima:2024cok} for technical details).

The agreement between the frequency- and the time-domain methods is overall very good, especially for the frequency of the mode. This is quantified in Table~\ref{tab:QNMs_BHs} for few selected cases and the first three tones.

\begin{table*}[h!]
    \centering
    \begin{tabular}{|c|c|c|c|c|c|c|}
     \hline
     \hline
  \multicolumn{3}{|c}{$r_S/r_B=$} \vline  & $3/2$ &  $6/5$ & $105/100$ & $101/100$ \\
    \hline
  \multirow{5}{*}{$l=1$}
    &
  \multirow{2}{*}{$p=0$}  
    & f-domain  & $0.475099 - \I 9.839 \times 10^{-2}$ 
                & $0.524964 - \I 9.099 \times 10^{-2}$ 
                & $0.560557 - \I 8.253 \times 10^{-2}$ 
                & $0.571244 - \I 7.323 \times 10^{-2}$ \\
   \cline{3-7}
    &
    & t-domain  & $0.475190 - \I 9.838 \times 10^{-2}$ 
                & $0.525453 - \I 9.027 \times 10^{-2}$ 
                & $0.560590 - \I 8.253 \times 10^{-2}$ 
                & $0.571628 - \I 7.295 \times 10^{-2}$ \\
   \cline{2-7}
    &
  \multirow{2}{*}{$p=1$}  
    & f-domain  & $0.354988 - \I 9.783 \times 10^{-2}$ 
                & $0.390223 - \I 9.481 \times 10^{-2}$
                & $0.420558 - \I 8.874 \times 10^{-2}$ 
                & $0.435112 - \I 8.038 \times 10^{-2}$ \\ 
   \cline{3-7}
    &
    & t-domain  & $0.354371 - \I 9.776 \times 10^{-2}$ 
                & $0.390621 - \I 9.238 \times 10^{-2}$
                & $0.421263 - \I 8.647 \times 10^{-2}$ 
                & $0.436811 - \I 7.940 \times 10^{-2}$ \\
    \hline
    \end{tabular}
    \begin{tabular}{|c|c|c|c|c|c|c|}
     \hline
     \hline
  \multicolumn{3}{|c}{$r_S/r_B=$} \vline  & $3/2$ &  $6/5$ & $105/100$
  & $101/100$ \\
    \hline
  \multirow{7}{*}{$l=2$}
    &
    \multirow{2}{*}{$p=0$}  
    & f-domain  & $0.746666 -\I 9.787 \times 10^{-2}$ 
                & $0.813286 -\I 9.169 \times 10^{-2}$
                & $0.860856 -\I 8.465 \times 10^{-2}$
                & $0.875909 -\I 8.194 \times 10^{-2}$
    \\
   \cline{3-7}
    &
    & t-domain & $0.746698 -\I 9.754 \times 10^{-2}$  
               & $0.812986 -\I 9.114 \times 10^{-2}$
               & $0.860790 -\I 8.451 \times 10^{-2}$
               & $0.875271 -\I 8.151 \times 10^{-2}$
    \\
   \cline{2-7}
   &
  \multirow{2}{*}{$p=1$}  
    & f-domain  & $0.623768 -\I 9.629 \times 10^{-2}$ 
                & $0.670552 -\I 9.213 \times 10^{-2}$
                & $0.706721 -\I 8.691 \times 10^{-2}$
                & $0.719101 -\I 8.441 \times 10^{-2}$   
    \\
   \cline{3-7}
    &
    & t-domain  & $0.622475 -\I 9.716 \times 10^{-2}$ 
                & $0.670021 -\I 9.250 \times 10^{-2}$
                & $0.706936 -\I 8.659 \times 10^{-2}$
                & $0.717444 -\I 8.551 \times 10^{-2}$ 
    \\
   \cline{2-7}
    &
  \multirow{2}{*}{$p=2$}  
    & f-domain  & $0.441539 -\I 8.722 \times 10^{-2}$ 
                & $0.458477 -\I 8.424 \times 10^{-2}$ 
                & $0.470346 -\I 8.122 \times 10^{-2}$
                & $0.473974 -\I 8.009 \times 10^{-2}$ 
    \\
   \cline{3-7}
    &
    & t-domain  & $0.441655 -\I 8.832 \times 10^{-2}$ 
                & $0.458697 -\I 8.449 \times 10^{-2}$ 
                & $0.470706 -\I 8.112 \times 10^{-2}$
                & $0.473640 -\I 8.005 \times 10^{-2}$  
    \\
    \hline
    \hline
    \end{tabular}
     \caption{Type-II QNMs of magnetized BH with $e Q_m/M \approx \left\lbrace 1.061\,, 1.116 \,, 1.153 \right\rbrace$ for $l=1$ (top table) and $l=2$ (bottom table).
     Modes with $p=0,1$ correspond to scalar-driven and electromagnetic-driven perturbations, whereas $p=2$ denotes gravitational-driven perturbations (which are not dynamical for $l=1$). 
     }\label{tab:QNMs_BHs}
\end{table*}

\begin{figure*}[th]
  \centering
  \includegraphics[width=1\textwidth]{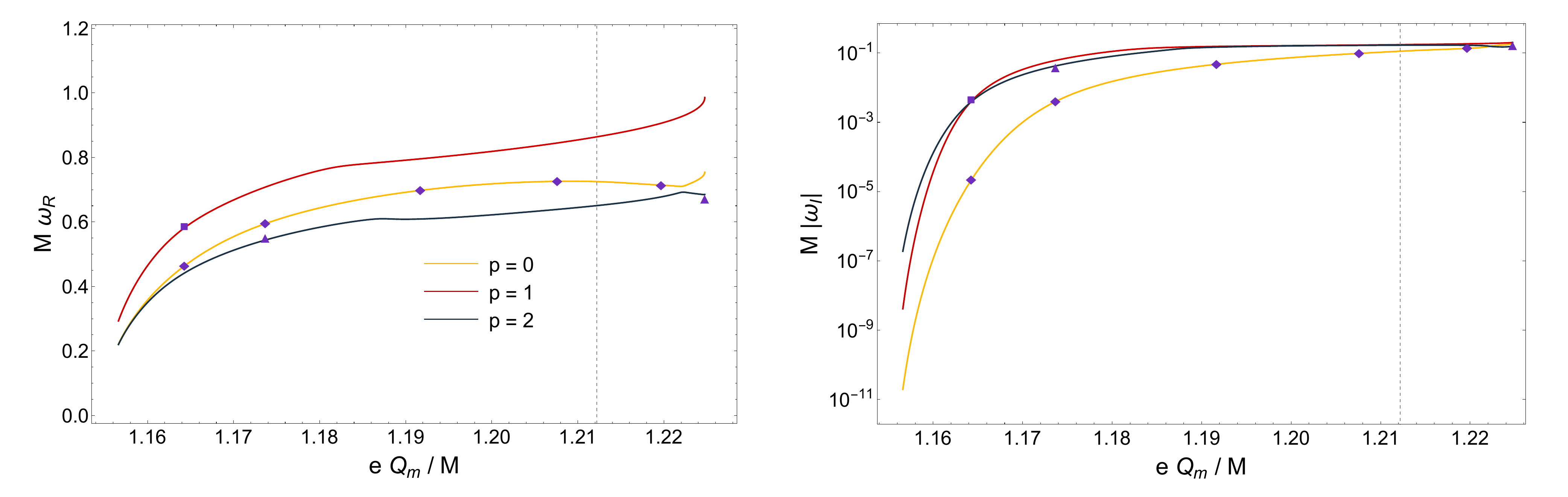}
  \includegraphics[width=1\textwidth]{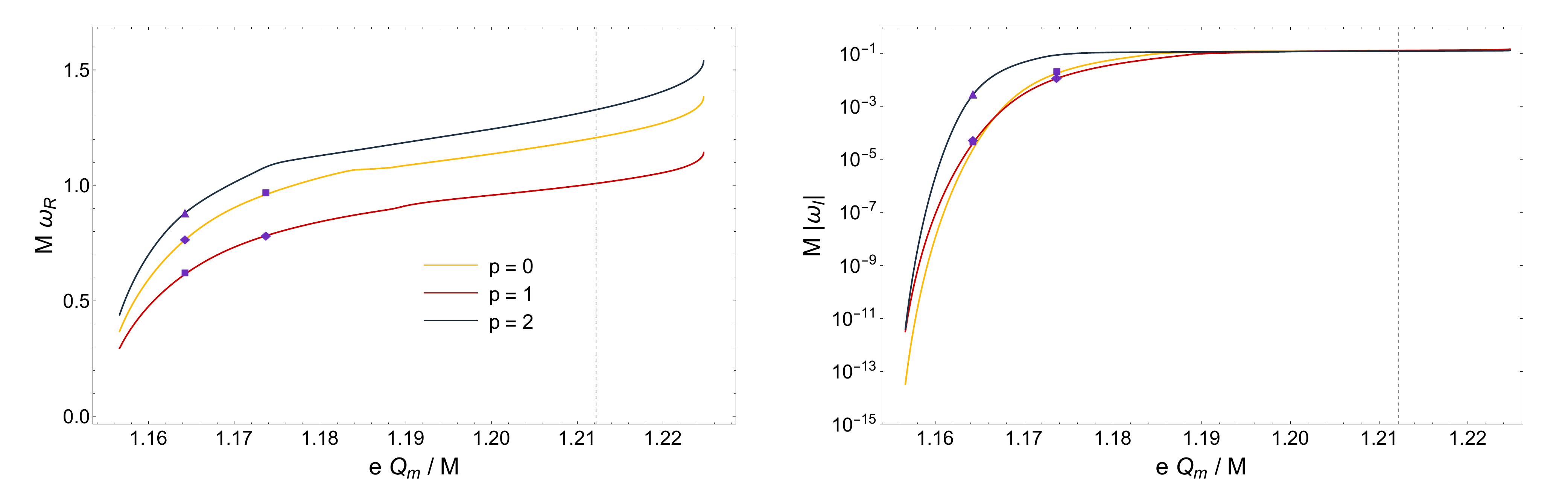}
  \caption{The first $p=0,1,2$ QNMs of a TS as a function of the magnetic charge for Type-II, $l=1$ (top panels) and $l=2$ (bottom panels) perturbations. The left (right) panels show the real (imaginary) part of the mode. The vertical dashed line denotes the transition between second-kind and first-kind TS.
  The markers corresponds to the modes extracted from the Fourier transform of the time-domain signal. Note that the mode hierarchy (labeled by $p$) depends on the charge, since there are crossings in the imaginary part.}\label{fig:QNMs_TypeII_l1l2_TS}
\end{figure*}

Let us now summarize the results for the Type-II QNMs of TSs. We show the most relevant modes for $l=1,2$ in Fig.~\ref{fig:QNMs_TypeII_l1l2_TS}.
In this case we show different modes ($p=0,1,2$), since they would be relevant to accurately fit the time-domain response, as we shall later discuss. Note that the mode hierarchy depends on the charge, since there are crossings in the imaginary part. We classify $p=0,1,2,...$ modes referring to the region near the BH transition (i.e., smallest value of $e Q_m/M$).
Also in this case we present a comparison between the QNMs computed in the frequency domain and those extracted from the time response (see Table~\ref{tab:QNMs_TSs}), showing very good agreement.

As already noted for the Type-I sector~\cite{Dima:2024cok}, the QNM spectrum of a TS smoothly interpolates from BH-like modes for first-kind TSs, to long-lived modes as the TS becomes of the second kind.

\begin{table*}[h!] 
    \centering
    \begin{tabular}{|c|c|c|c|c|c|c|}
     \hline
     \hline
  \multicolumn{3}{|c}{$r_S/r_B=$} \vline  & $7/10$ & $9/10$ & $19/20$ & $99/100$ \\
    \hline
  \multirow{10}{*}{$l=1$}
    &
  \multirow{2}{*}{$p=0$}  
    & f-domain & $0.725735 - \I 9.739 \times 10^{-2}$  
               & $0.594653 - \I 3.979 \times 10^{-3}$
               & $0.463193 - \I 2.163 \times 10^{-5}$ 
               & $0.221521 - \I 1.967 \times 10^{-11}$ \\
   \cline{3-7}
    &
    & t-domain  & $0.725753 - \I 9.735 \times 10^{-2}$ 
                & $0.594636 - \I 3.978 \times 10^{-3}$ 
                & $0.463183 - \I 2.160 \times 10^{-5}$ 
                & $0.221438 - \I 2.110 \times 10^{-11}$ \\
   \cline{2-7}
    &
  \multirow{2}{*}{$p=1$} 
    & f-domain & - 
               & $0.543775 - \I 4.233 \times 10^{-2}$  
               & $0.581303 - \I 4.013 \times 10^{-3}$ 
               & $0.293106 - \I 4.147 \times 10^{-9}$ \\ 
    \cline{3-7}
    &
    & t-domain  & $-$
                & $0.549296 - \I 3.719 \times 10^{-2}$
                & $0.581265 - \I 4.006 \times 10^{-3}$ 
                & $0.292507 - \I 4.089 \times 10^{-9}$ \\
    \cline{2-7}
    &
  \multirow{2}{*}{$p=2$} 
    & f-domain & - 
               & $0.708010 - \I 6.095 \times 10^{-2}$  
               & $0.441270 - \I 3.926 \times 10^{-3}$
               & $0.220050 - \I 1.896 \times 10^{-7}$ \\ 
    \cline{3-7}
    &
    & t-domain  & $-$
                & $0.709243 - \I 5.997 \times 10^{-2}$
                & $0.441245 - \I 3.910 \times 10^{-3}$ 
                & $0.220049 - \I 1.875 \times 10^{-7}$ \\
   \cline{2-7}
    &
  \multirow{2}{*}{$p=3$} 
    & f-domain & - 
               & -  
               & $0.681979 - \I 3.779 \times 10^{-2}$ 
               & $0.362912 - \I 2.424 \times 10^{-7}$ \\ 
    \cline{3-7}
    &
    & t-domain  & $-$
                & $-$
                & $0.682161 -\I 3.173 \times 10^{-2}$ 
                & $0.362168 - \I 2.329 \times 10^{-7}$ \\
    \hline
    \end{tabular}
    \begin{tabular}{|c|c|c|c|c|c|c|}
     \hline
     \hline
  \multicolumn{3}{|c}{$r_S/r_B=$} \vline  & $7/10$ & $9/10$ & $19/20$ & $99/100$ \\
    \hline
  \multirow{10}{*}{$l=2$}
    &
  \multirow{2}{*}{$p=0$}  
    & f-domain & $0.569028 - \I 4.955 \times 10^{-2}$  
               & $0.834579 - \I 1.619 \times 10^{-4}$ 
               & $0.630961 - \I 2.895 \times 10^{-8}$ 
               & $0.368629 - \I 3.221 \times 10^{-14}$ \\
   \cline{3-7}
    &
    & t-domain & $0.569024 - \I 4.981 \times 10^{-2}$ 
               & $0.834486 - \I 1.865 \times 10^{-4}$ 
               & $0.630844 - \I 3.245 \times 10^{-8}$
               & $0.368733 - \I (*)$  \\
  \cline{2-7}
    &
  \multirow{2}{*}{$p=1$} 
    & f-domain & $1.085261 - \I 7.599 \times 10^{-2}$ 
               & $0.422771 - \I 6.856 \times 10^{-4}$  
               & $0.317783 - \I 5.510 \times 10^{-6}$ 
               & $0.295274 - \I 3.206 \times 10^{-12}$ \\ 
  \cline{3-7}
    &
    & t-domain  & $1.085059 - \I 7.557 \times 10^{-2}$
                & $0.424131 - \I 7.271 \times 10^{-4}$ 
                & $0.317827 - \I 5.551 \times 10^{-6}$ 
                & $0.295660 - \I  (*) $ \\
   \cline{2-7}
    &
  \multirow{2}{*}{$p=2$} 
    & f-domain & - 
               & $0.782183 - \I 1.144 \times 10^{-2}$  
               & $0.764893 - \I 2.534 \times 10^{-5}$ 
               & $0.439992 - \I 3.944 \times 10^{-12}$ \\ 
    \cline{3-7}
    &
    & t-domain  & -
                & $0.782285 - \I 1.146 \times 10^{-2}$ 
                & $0.765489 - \I 3.821 \times 10^{-5}$ 
                & $0.440391 - \I (*) $ \\
   \cline{2-7}
    &
  \multirow{2}{*}{$p=3$} 
    & f-domain & - 
               & $0.960878 - \I 1.792 \times 10^{-2}$  
               & $0.616095 - \I 4.239 \times 10^{-5}$ 
               & $0.148446 - \I 6.337 \times 10^{-11}$ \\ 
    \cline{3-7}
    &
    & t-domain  & -
                & $0.961672 - \I 1.825 \times 10^{-2}$ 
                & $0.616099 - \I 4.103 \times 10^{-5}$ 
                & $0.148477 - \I 6.262 \times 10^{-11}$ \\
    \hline
    \hline
    \end{tabular}
     \caption{Same as in Table~\ref{tab:QNMs_BHs} but for TSs with $e Q_m/M \approx \left\lbrace 1.208\,, 1.174\,, 1.164\,, 1.157 \right\rbrace$ (equivalently, $r_S/r_B= \left\lbrace 0.70\,, 0.90\,, 0.95\,, 0.99 \right\rbrace$). The asterisk indicates QNMs with a characteristic damping time that is too large for the spectral analysis to retrieve an accurate fit thereof.}\label{tab:QNMs_TSs}
\end{table*}

\begin{figure*}[th]
  \centering
  \includegraphics[width=0.5\textwidth]{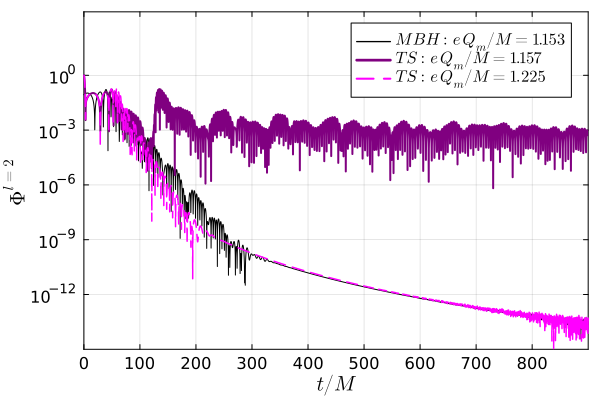}
  \includegraphics[width=0.5\textwidth]{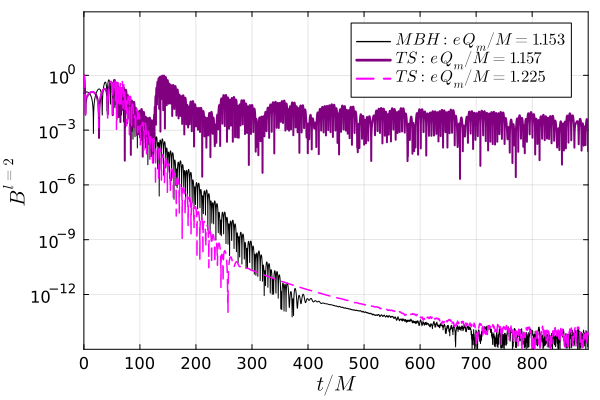}
  \includegraphics[width=0.5\textwidth]{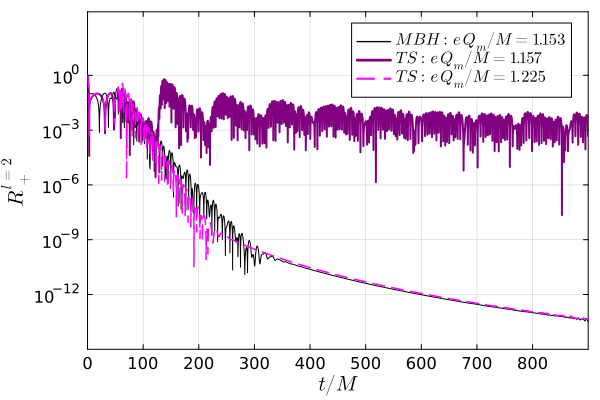}
  \caption{Comparison between the linear response of a nearly-extremal magnetized BH (black), a second-kind TS with same mass and similar charge-to-mass ratio (purple), and a first-kind TS (magenta) to $l=2$ even-parity (Type-II) perturbations. Top, middle, and bottom panels refer to scalar-driven, electromagnetic-driven, and gravitational-driven perturbations, respectively.
  } \label{fig:Phi_RingdownVSEchoes}
\end{figure*}

Because of these long-lived modes, it is natural to ask whether there exist nonradial unstable modes
(which, within our conventions, are defined by a QNM with positive imaginary part).
We have scanned the parameter space of TSs searching for such unstable modes and did not found any. This is also confirmed by the time domain response, which is always decaying in time. Some representative examples are shown in Fig.~\ref{fig:Phi_RingdownVSEchoes} for the scalar-driven, electromagnetic-driven, and gravitational-driven modes of a magnetized BH, second-kind TSs with similar charge-to-mass ratio, and a first-kind TS.
As discussed in Ref.~\cite{Dima:2024cok} for Type-I perturbations, also in this case we observe that the ringdown of a second-kind TS can be initially very similar to that of a BH with similar charge-to-mass ratio, whereas differences appear at late times in the form of (slowly decaying) echoes arising from the absence of a horizon in the TS spacetime.
On the other hand, less compact (first-kind) TSs display a standard exponentially suppressed ringdown, but with different modes compared to the BH case.

Finally, we also note that TSs have the same late-time, power-law tail as BHs~\cite{Price:1972pw}, regardless of the charge-to-mass ratio. While the scaling in time of the tail is due to low-frequency backscattering off the potential and is indeed universal and independent of the nature of the object, it is interesting to note that the amplitude of the tails seems also universal. We have checked that the latter depends only on the amplitude of the initial data and not on the underlying solutions.

\section{Conclusions}

We studied nonradial perturbations of magnetized BHs and TSs with zero momentum along the fifth dimension, providing strong numerical evidence for the linear stability of these solutions in a region not affected by the known Gregory-Laflamme-like instability for radial perturbations.
It would be interesting to explore if the stability results extend also for nonradial perturbations with nonzero Kaluza-Klein momentum. This would require a generalization of the perturbation ansatz, since more perturbations of the five-dimensional fields would be allowed in the presence of Kaluza-Klein momentum.

A natural follow-up consists in verifying the stability under nonlinear perturbations, especially for TSs of the second kind, which feature a stable photon sphere. Indeed, there are arguments suggesting that ultracompact objects might be unstable at the nonlinear level~\cite{Cardoso:2014sna,Cunha:2017qtt}.
This is due to the slow (possibly logarithmic) decay in time of their characteristic modes, as discussed for microstate geometries~\cite{Eperon:2016cdd} and for other ultracompact objects~\cite{Cardoso:2014sna,Keir:2014oka}.
TSs provide a well-defined model in which the nonlinear evolution of the perturbations can be possibly studied in a relatively simple setting.

Finally, our results can be also useful to study the gravitational-wave emission from a point particle orbiting a TS, as recently studied for a test scalar field~\cite{Bianchi:2024rod}.

\begin{acknowledgments}
We are indebted to Nicola Franchini for enlightening discussions, comments on the draft, and for spotting an error in the initial approach to the special case $l=1$. 
This work is partially supported by the MUR PRIN Grant 2020KR4KN2 ``String Theory as a bridge between Gauge Theories and Quantum Gravity'', by the FARE programme (GW-NEXT, CUP:~B84I20000100001), and by the INFN TEONGRAV initiative.
Some numerical computations have been performed at the Vera cluster supported by the Italian Ministry for Research and by Sapienza University of Rome.
\end{acknowledgments}

\newpage

\appendix

\section{Regge-Wheeler-Zerilli ansatz for 5D linear perturbations} \label{app:RW}

We employ an extended Regge-Wheeler ansatz for 5D perturbations (see \cite{Pereniguez:2023wxf} for a discussion on how to fix the gauge starting from the most general ansatz), to decouple the angular dependence:

\begin{widetext}
    
\begin{align}
h_{AB}^{\rm even} & =\sum_{l,m}\left(
\begin{array}{ccccc}
~\label{eq:rw_ansatz_even_gab}
  f_S H_0(t,y,r) & H_4(t,y,r) &  H_1(t,y,r) & 0 & 0 \\
   * & f_B H_3(t,y,r) & H_5(t,y,r) & 0 & 0 \\
   * & * & (f_Sf_B)^{-1} H_2(t,y,r) & 0 & 0 \\
   0 & 0 & 0 & r^2  K(t,y,r) & 0 \\
   0 & 0 & 0 & 0 & r^2  \sin\theta^2 K(t,y,r)
\end{array}
\right)  Y_{lm}(\theta,\phi)\,,
\\
h_{AB}^{\rm odd} & =\sum_{l,m}\left(
\begin{array}{ccccc}
~\label{eq:rw_ansatz_odd_gab}
    0 & 0 & 0 & -h_0(t,y,r)/\sin\theta \partial_\phi & h_0(t,y,r) \sin\theta \partial_\theta \\
    0 & 0 & 0 & -h_2(t,y,r)/\sin\theta \partial_\phi & h_2(t,y,r) \sin\theta \partial_\theta \\
    0 & 0 & 0 & -h_1(t,y,r)/\sin\theta \partial_\phi & h_1(t,y,r) \sin\theta \partial_\theta \\
    * & * & * & 0 & 0 \\
    * & * & * & 0 & 0
\end{array}
\right) Y_{lm}(\theta,\phi)\,,
\\
  f_{AB}^{\rm even} & =\sum_{l,m}\left(
  \begin{array}{ccccc}~\label{eq:rw_ansatz_even_em}
     0             & f_{ty}^+(t,y,r)      & f_{tr}^+(t,y,r) & f_{t\theta}^+(t,y,r) \partial_\theta & f_{t\theta}^+(t,y,r) \partial_\phi \\
    \bdot &             0      & f_{yr}^+(t,y,r) & f_{y\theta}^+(t,y,r) \partial_\theta & f_{y\theta}^+(t,y,r) \partial_\phi \\
    \bdot & \bdot     &            0  & f_{r\theta}^+(t,y,r) \partial_\theta & f_{r\theta}^+(t,y,r) \partial_\phi \\
    \bdot & \bdot  & \bdot & 0 & 0  \\
    \bdot & \bdot  & \bdot & 0 & 0
  \end{array}
  \right) Y_{lm}(\theta,\phi)\,,
  \\
  f_{AB}^{\rm odd} & =\sum_{l,m}\left(
  \begin{array}{ccccc}~\label{eq:rw_ansatz_odd_em}
    0 & 0 & 0 & f_{t\theta}^-(t,y,r)/\sin\theta \partial_\phi & -f_{t\theta}^-(t,y,r) \sin\theta \partial_\theta \\
    0 & 0 & 0 & f_{y\theta}^-(t,y,r)/\sin\theta \partial_\phi & -f_{y\theta}^-(t,y,r) \sin\theta \partial_\theta \\
    0 & 0 & 0 & f_{r\theta}^-(t,y,r)/\sin\theta \partial_\phi & -f_{r\theta}^-(t,y,r) \sin\theta \partial_\theta \\
    \bdot & \bdot & \bdot & 0 & l(l+1) f_{\theta\phi}^-(t,y,r) \sin\theta  \\
    \bdot & \bdot & \bdot & \bdot & 0
  \end{array}
  \right) Y_{lm}(\theta,\phi)
  \,.
\end{align}

\end{widetext}
where the uppercase Latin indices span the coordinates $A\,,B=\lbrace t\,, y \,, r \,, \theta \,, \phi \rbrace$. The asterisks, $*$, (respectively, dots, $\cdot$) indicate symmetry (respectively, antisymmetry) with respect to the corresponding upper-half matrix elements.

Furthermore, because the background possesses two isometries in the $t$ and $y$ directions,
 we can further decouple the time and fifth-dimension dependence by:
\begin{equation}\label{eq:Fourier}
    A(t,y,r)=e^{-\I \omega t}e^{\I \sigma y} a(r)
\end{equation}
where $A$ is the generic perturbation field. In this work, we restrict to consider the scenario in which the fifth dimension is compactified, our compact objects are macroscopic and effectively 4D and, thus, excitations in the fifth dimension are negligible, $\sigma=0$ (see the discussion in \cite{Dima:2024cok}).

\section{Evolution equations for linear perturbations}
\label{app:deteqs}

By applying the ans{\"a}tze \eqref{eq:rw_ansatz_even_gab} and \eqref{eq:Fourier} to the field equations, and keeping only linear terms, one obtains a set of first- and second-order partial differential equations that can be further recast to obtain the evolution equations for the proper independent dynamical degrees of freedom. Because of the magnetic charge of the background, the perturbations will be coupled together in the equations according to two groups: Type-I perturbations ($h_{AB}^{\mathrm{odd}}$ and $f_{AB}^{\mathrm{even}}$) and Type-II perturbations ($h_{AB}^{\mathrm{even}}$ and $f_{AB}^{\mathrm{odd}}$). The first group (as well as the special case $l=0$ from the second) have been studied and discussed in \cite{Dima:2024cok}. Here, we shall focus on the solution and discussion of Type-II perturbations in the two cases, $l=1$ and $l\geq 2$.

\subsection{Type-II, $l\geq 2$}

The derivation of the system of equations~\eqref{eq:TypeII_l2} valid for the generic case $l\geq 2$ involves some cumbersome manipulations, thus here we will summarize the main idea and discuss only the key steps.

First, we proceed by taking the limit $\sigma\rightarrow 0$, which is the most interesting one for macroscopic objects. Then, we can eliminate $H_2$ from all equations by employing the $(_\theta^\phi)$ component of the Einstein equations. 
Next, by combining the $(_t^r)$, $(_r^\phi)$ and $(_r^r)$ components, an equation can be obtained that relates $H_0$ to an algebraic combination of $H_1$, $K$, $f_{\theta\phi}^-$, $H_3$ and first derivatives thereof. This relation can be inverted to eliminate $H_0$ from the system of equations just as we did for $H_2$.

With the elimination of $H_0$, $H_2$, we can construct a
second-order equation for $H_3$ via a linear combination of $(_y^y)$, $(_\theta^\theta)$ components of the linearized Einstein equations, in addition to the second-order equation for $f_{\theta\phi}^-$, which is provided by the $\theta$-component of the linearized Maxwell equation. 
In addition, the linearized Einstein equations are reduced to a system of two first-order equations for $H_1$ and $K$.
We then proceed to generalize Zerilli's procedure to reduce the system of two fist-order equations to a single second-order equation for an auxiliary variable that encodes the gravitational degree of freedom. To do so, we introduce two auxiliary variables $\hPsi = \left( \hK \,, \hH \right)$ linked to $\Psi=\left( K \,, H_1 \right)$ via $\Psi = \mathbf{F} \hPsi$, and such that the original system of first-order equations
\begin{align}
    \frac{d\Psi}{dr} & =\mathbf{A}\cdot \Psi + \mathcal{S}
\end{align}
reduces to 
\begin{align}
    \frac{d\hPsi}{dr_*} & =\mathbf{B}\cdot \hPsi + \hat{\mathcal{S}}
    \,,
\end{align}
where we introduced the generic coordinate $dr_*=n(r)^{-1}dr$.
The matrix elements of $\mathbf{A}$ are known, the elements of $\mathbf{F}$ are unknown and we fix the form of $\mathbf{B}$ to be
\begin{align}
\mathbf{B} & =
    \left(
    \begin{array}{cc}
       q(r) & 1  \\
       V_0(r) - \omega^2 & p(r)
    \end{array}
    \right)
\end{align}
where $p(r)$, $q(r)$ and $V_0(r)$ are also free to be fixed.

One can notice that the equations above imply the following system:
\begin{align}~\label{eq:sistemone}
    n(r) \mathbf{F}^{-1}\left((\mathbf{A}\mathbf{F}-\frac{d \mathbf{F}}{dr}) \right) - \mathbf{B} = 0
\end{align}
which can be solved to fix $\mathbf{F}$ and $\mathbf{B}$.
This step is as crucial as delicate. We first fix $n(r):=(\sqrt{f_B}f_S)^{-1}$, in such a way that the coordinate $r_*$ is the appropriate tortoise one for our background \footnote{For more details on the tortoise coordinate and its adaptation to both magnetized BH and TS backgrounds, see discussion in~\cite{Dima:2024cok}.}. 

The result of this procedure can be summed up in the relations:
\begin{align}
    K & = -\frac{(1+3f_B)}{8r f_B\zeta_B} \partial_r\hK 
        - \frac{3+4\lambda+3f_B(f_S-1)-3f_S}{8r^2f_Bf_S\zeta_B}\hK
        \nonumber\\
        &
        - \frac{P\k5^2(1+3f_B)^2f_S}{4r^2f_B\zeta_B}f_{\theta\phi}^-
        + \frac{r(1+3f_B)f_S}{4\zeta_B} \partial_r H_3
        \nonumber\\
        &
        + \frac{(1+3f_B)(2f_S+f_B(3f_S-1))}{8f_B\zeta_B} H_3
    \\
    H_1 & = -\frac{4r}{f_S(1+3f_B)}K - \frac{1}{rf_Bf_S^2(1+3f_B)}\hK
\end{align}
where $2\lambda=(l-1)(l+2)$ and $8f_B\zeta_B = f_B^2(6-15f_S)+2f_B(9+8\lambda-6f_S)+3f_S$.
As final steps, we introduce $\Rp := \hK - 2r^2f_Bf_S H_3$, $\Phi := H_3/f_B$ and $\B := f_{\theta\phi}^-/\k5$, to help us simplify the equations and obtain the final system~\eqref{eq:TypeII_l2} with coefficients defined as:
\begin{widetext}
    \begin{align}
     \tilde{U}_{ij} & = -\delta_{i1}\delta_{j3} \frac{(1-f_B)^2f_S^3}{4\zeta_B}
    \\
    \tilde{W}_{11} & = -\frac{f_S}{8r\zeta_B}
    \left(\frac{}{}(15 + 16 \lambda - 9 f_S) f_S - 
 2 f_B (9 + 8 \lambda + (9 + 16 \lambda) f_S) + 
 3 f_B^2 (-2 + f_S + 11 f_S^2)\frac{}{}\right)
    \\
    \tilde{W}_{12}  & = -\frac{P\k5(1-f_B)f_S^2}{4r^3\zeta_B}
    \left(\frac{}{}-2 f_B (9 + 8 \lambda - 8 f_S) - f_S + f_B^2 (-6 + 9 f_S)\frac{}{}\right)
    \\
    \tilde{W}_{13}  & = - \frac{(1-f_B)f_Bf_S^2}{64rf_B^2\zeta_B^2}
    \left(\frac{}{} f_S (-33 - 32\lambda + 39 f_S) + 
 f_B (6 (9 + 8\lambda) + (111 + 112\lambda) f_S - 
    87 f_S^2) 
 \right.
 \nonumber\\
 &\left.
 + 
 f_B^2 (108 + 80\lambda - (267 + 208\lambda) f_S - 
    15 f_S^2) + 15 f_B^3 (2 - 13 f_S + 17 f_S^2)\frac{}{}\right)
    \\
\tilde{V}_{11}& = \frac{f_S}{64r^2f_B^2\zeta_B^2}
    \left(\frac{}{} 9 (7 + 6 \lambda - 3 f_S) f_S^2 + 
 f_B f_S (189 + 312 \lambda + 128 \lambda^2 - 
    15 (21 + 16 \lambda) f_S + 72 f_S^2) 
 \right.
 \nonumber\\
 &\left.
 - 
 3 f_B^4 (66 + 72 \lambda - 6 (31 + 28 \lambda) f_S + 
    7 (27 + 2 \lambda) f_S^2 - 201 f_S^3) + 
 2 f_B^2 ((7 + 4 \lambda) (9 + 8 \lambda)^2 - (927 + 
       1404 \lambda + 512 \lambda^2) f_S 
 \right.
 \nonumber\\
 &\left.
 + 
    6 (96 + 71 \lambda) f_S^2 - 72 f_S^3) + 
 9 f_B^5 (-6 + 31 f_S - 41 f_S^2 + 10 f_S^3) - 
 2 f_B^3 (-15 (9 + 8 \lambda) + (738 + 924 \lambda + 
       320 \lambda^2) f_S 
 \right.
 \nonumber\\
 &\left.
 - 66 (9 + 4 \lambda) f_S^2 + 
    297 f_S^3)\frac{}{}\right)
    \\
\tilde{V}_{12}& = \frac{P\kappa_5f_Bf_S}{96r^4f_B^2\zeta_B^2}
    \left(\frac{}{} 3 (45 + 44 \lambda - 51 f_S) f_S^2 + 
 27 f_B^5 f_S (10 - 41 f_S + 25 f_S^2) + 
 3 f_B f_S (-6 (9 + 8 \lambda) - 5 (33 + 32 \lambda) f_S + 
    189 f_S^2)    
 \right.
 \nonumber\\
 &\left.
 + 
 6 f_B^3 (-16 (9 + 17 \lambda + 8 \lambda^2) + (438 + 
       856 \lambda + 384 \lambda^2) f_S 
       - (189 + 400 \lambda) f_S^2 - 63 f_S^3) - 
 9 f_B^4 (16 (1 + \lambda) - 8 (23 + 22 \lambda) f_S     
 \right.
 \nonumber\\
 &\left.
+ 
    3 (71 + 52 \lambda) f_S^2 - 51 f_S^3) - 
 2 f_B^2 (8 (1 + \lambda) (9 + 8 \lambda)^2 - 
    12 (9 + 46 \lambda + 32 \lambda^2) f_S - 
    3 (369 + 308 \lambda) f_S^2 + 585 f_S^3)\frac{}{}\right)
    \\
\tilde{V}_{13}& = -\frac{(1-f_B)f_S}{32r^2f_B^2\zeta_B^2}
    \left(\frac{}{} 9 (1 + \lambda) f_S^2 - 
 f_B f_S (2 (9 + 17 \lambda + 8 \lambda^2) + (12 + 13 \lambda) f_S    + 9 f_S^2) + 3 f_B^4 f_S (10 - 53 f_S + 55 f_S^2)      
 \right.
 \nonumber\\
 &\left.
+
  f_B^2 (8 (9 + 17 \lambda + 8 \lambda^2) - 
    6 (15 + 22 \lambda + 8 \lambda^2) f_S + (78 + 63 \lambda) f_S^2 + 27 f_S^3) - 
 f_B^3 (-24 (1 + \lambda) + 2 (9 + 13 \lambda) f_S     
 \right.
 \nonumber\\
 &\left.
 + 
    3 (4 + 9 \lambda) f_S^2 + 87 f_S^3)\frac{}{}\right)
\end{align}

\end{widetext}

\begin{widetext}
    
\begin{align}
    \tilde{W}_{21} & = 0
    \\
    \tilde{W}_{22}  & = \frac{f_S}{r}
    \left(\frac{}{}f_B+f_S-2f_Bf_S\frac{}{}\right)
    \\
    \tilde{W}_{23}  & = - \frac{P\k5(1+3f_B)f_S^2}{8rf_B\zeta_B}
    \\
\tilde{V}_{21}& = -\frac{P\k5f_S}{8r^2f_B\zeta_B}
    \left(\frac{}{} 9 + 8 \lambda + 3 f_B (1 - f_S) - 9 f_S\frac{}{}\right)
    \\
\tilde{V}_{22}& = \frac{f_S}{8r^2}
    \left(\frac{}{} 16(1+\lambda)
+ \frac{3(1-f_B)(1-f_S)(1+3f_B)^2f_S}{f_B\zeta_B}
\frac{}{}\right)
    \\
\tilde{V}_{23}& = -\frac{P\k5f_S}{4r^2f_B\zeta_B}
    \left(\frac{}{} 2(1+\lambda)-(1-3f_B)f_S\frac{}{}\right)
\end{align}

\end{widetext}

\begin{widetext}
    
\begin{align}
    \tilde{W}_{31} & = 0
    \\
    \tilde{W}_{32}  & = 0
    \\
    \tilde{W}_{33}  & = - \frac{f_Bf_S}{8rf_B\zeta_B}
    \left(\frac{}{}(21 + 16 \lambda - 27 f_S) f_S - 
 2 f_B (9 + 8 \lambda - 3 f_S) (1 + 2 f_S) + 
 f_B^2 (-6 - 39 f_S + 87 f_S^2)\frac{}{}\right)
    \\
\tilde{V}_{31}& = -\frac{3(1-f_B)f_S}{8r^2f_B\zeta_B}
    \left(\frac{}{} f_S (-9 - 8 \lambda + 3 f_S) + 
 f_B^2 (6 - 9 f_S + 9 f_S^2) + 
 2 f_B (9 + 8 \lambda - (27 + 20 \lambda) f_S + 18 f_S^2) \frac{}{}\right)
    \\
\tilde{V}_{32}& = \frac{P\k5f_Bf_S}{4r^4f_B\zeta_B}
    \left(\frac{}{} f_S (-5 (9 + 8 \lambda) + 33 f_S) + 
 9 f_B^3 (2 + f_S + f_S^2) - 
 f_B (2 (63 + 128 \lambda + 64 \lambda^2) - (99 + 
       112 \lambda) f_S + 9 f_S^2)      
 \right.
 \nonumber\\
 &\left.
+ 
 3 f_B^2 (4 - 3 (7 + 8 \lambda) f_S + 21 f_S^2)
\frac{}{}\right)
    \\
\tilde{V}_{33}& = \frac{f_S}{4r^2f_B\zeta_B}
    \left(\frac{}{} -3 (1 + \lambda) f_S + 
 f_B^2 (-36 - 38 \lambda + (57 + 39 \lambda) f_S - 
    42 f_S^2) + 
 f_B (30 + 46 \lambda + 
    16 \lambda^2 - (33 + 28 \lambda) f_S       
 \right.
 \nonumber\\
 &\left.
+ 12 f_S^2) + 
 3 f_B^3 (-2 - 7 f_S + 14 f_S^2) \frac{}{}\right)
\end{align}

\end{widetext}

\subsection{Type-II, $l=1$}

In the special case $l=1$ the Type-II perturbation fields are $\lbrace H_0, H_1, H_2, H_3, H_4, H_5, f_{t\theta}^-, f_{r\theta}^-, f_{\theta\phi}^-\rbrace$, as $K$ is not defined for $l<2$. In addition, by applying the Fourier decomposition \eqref{eq:Fourier} and assuming $\sigma=0$, one notices that $H_4\,, H_5\,, f_{y\theta}^-$ decouple from the rest and can, thus, be neglected. Furthermore, because $f_{AB}$ are perturbations of a Maxwell tensor derived from a vector potential, one can check that $\lbrace f_{t\theta}^-\,, f_{r\theta}^- \,, f_{\theta\phi}^- \rbrace$ are related via $f_{t\theta}^-=-\I \omega f_{\theta\phi}^-$ and $f_{r\theta}^-=\partial_r f_{\theta\phi}^-$. 
After some algebraic manipulations, one can show that $H_0$ and $H_2$ can be recast as functions of the remaining perturbation fields, by using the $(r,r)$ and $(t,r)$ components of the perturbed Einstein equations.
What remains is the group of variables $\lbrace H_1\,, H_3 \,, f_{\theta\phi}^- \rbrace$. The first variable satisfies a first order equation in $r$ given by the $(t,\phi)$ component of the Einstein equations; $H_3$ satisfies a second order equation constructed as an appropriate combination of the $(t,t)$, $(y,y)$ and $(\theta,\theta)$ components; and, finally, a second order equation for $f_{\theta\phi}^-$ is obtained from the perturbed Maxwell equations.

Now, we introduce two dynamical variables, $\Upsilon_i$ with $i=1\,,2$, and the constrained variable $\Lambda$ that relate to $\lbrace H_1\,,H_3\,,f_{\theta\phi}^-\rbrace$ via

\begin{align}
    H_1 & = \Lambda - \frac{2r}{f_S(1+3f_B)} H_3\,, \label{ansatz}
    \\
    H_3 & = 
    \left[\left(1-\frac{4P^2\kappa_5^2}{(3-f_S)(9r^2-2P^2\kappa_5^2-3r^2f_S)}\right)\frac{1}{rf_B^2}\Upsilon_1
     \right.
     \nonumber\\
     & \hspace{2em}\left.
     +\frac{2P\kappa_5}{f_B(9r^2-2P^2\kappa_5^2-3r^2f_S)}\Upsilon_2 
     \right.
     \nonumber\\
     & \hspace{2em}\left.
    -\frac{(1-f_B)(1+3f_B)f_S}{3rf_B(3-f_S+f_B(1-3f_S))}\Lambda\right]\,,
    \\
     f_{\theta\phi}^- & = 
    \kappa_5^{-1} \left[ \Upsilon_2
     - \left(\frac{2P\kappa_5}{rf_B(3-f_S)}\right) \Upsilon_1
     \right.
     \nonumber\\
     & \hspace{3.5em}\left.
     + \frac{P\kappa_5f_S(1+3f_B)}{3r(3-f_S+f_B(1-3f_S))}\Lambda\right]\,.
\end{align}

The introduction of the auxiliary variables $\lbrace \Lambda\,,\Upsilon_1\,,\Upsilon_2 \rbrace$ allows us to recast the second order equations for $\lbrace H_3\,,f_{\theta\phi}^-\rbrace$ (which featured source terms proportional to $\lbrace H_1,H_3\,,f_{\theta\phi}^-\rbrace$ and derivatives thereof) into a pair of coupled second-order equations that only contain $\lbrace\Upsilon_i\,,\partial_r\Upsilon_i\,,\partial_r^2\Upsilon_i\rbrace$, which we listed in Eqs.~\eqref{eq:TypeII_l1}.
Finally, the ansatz \eqref{ansatz} allows us to realize that $\Lambda$ is related to the remaining two variables via a differential constrained equation that closes the system:
\begin{align}
  &\partial_r\Lambda
  +\mu_0 \Lambda
  +\mu_1 \Upsilon_1
  +\mu_2 \Upsilon_2 = 0
\end{align}
where
\begin{align}
  \mu_0 & = \left(\frac{1}{r f_B f_S(1 + 3 f_B) (-3 + f_S + f_B (-1 + 3 f_S))}\cdot
     \right.
     \nonumber\\
     & \hspace{2em}
  \cdot \left(\frac{}{}-3 + 3 f_B^3 (-1 + f_S) + 4 f_S - f_S^2 + 
  f_B^2 (-13 
     \right.
     \nonumber\\
     & \hspace{3em}\left.\left.
 + 30 f_S  - 9 f_S^2) + 
  f_B (-13 + 11 f_S - 6 f_S^2)\frac{}{}\right)\frac{}{}\right)\,,
  \\
\mu_1 & =
  \frac{1}{r^3f_Bf_S^2}\left(\frac{4P^2\kappa_5^2f_S}{3-f_S}-\frac{3 r^2(3 + f_B)(1-f_S)}{(1 + 3 f_B)^2}\cdot
     \right.
     \nonumber\\
     & \hspace{2em}\left.
     \cdot\left(1+ \frac{2}{3-f_S}-\frac{6r^2}{9r^2-2P^2\kappa_5^2-3r^2f_S}\right)\right)\,,
  \\
  \mu_2 & = - \frac{2P \kappa_5}{f_S^2} \left(\frac{f_S}{r^2} + \frac{3(3+f_B)(1-f_S)}{(1+3f_B)^2(9r^2-2P^2\kappa_5^2-3r^2f_S)} \right)\,.
\end{align}

For completeness, here we list the full expressions of the coefficients in Eqs.~\eqref{eq:TypeII_l1}:

\begin{widetext}
    
\begin{align}
    W_{11} & = \frac{f_S}{r(1+3f_B)(3-f_S)(2+f_B(1-f_S))^2} \cdot
        \nonumber\\
        &\hspace{1em}\cdot \left( -12 (-3 + f_S) f_S + 
 3 f_B^4 (-1 + f_S)^2 (3 + f_S + 2 f_S^2) + 
 4 f_B (3 + 14 f_S - 21 f_S^2 + 4 f_S^3) 
         \right.
         \nonumber\\
        &
        \hspace{2em}\left.+ 
 f_B^2 (48 - 119 f_S + 23 f_S^2 + 27 f_S^3 - 3 f_S^4) + 
 f_B^3 (39 - 174 f_S + 208 f_S^2 - 74 f_S^3 + f_S^4)\right)
    \\
    V_{11} & = \frac{f_S}{r^2(1+3f_B)^2(3-f_S)^2(2+f_B(1-f_S))^3(3-f_S+f_B(1-3f_S))^2} \cdot
                \nonumber\\
                &\hspace{1em}\cdot\left(\frac{}{} 12 (-3 + f_S)^3 f_S (-21 + f_S^2) + 
 54 f_B^9 (-1 + f_S)^4 (-3 - 2 f_S + f_S^2) - 
 4 f_B (-3 + f_S)^2 (-63 - 876 f_S + 1315 f_S^2 
                \right.
                \nonumber\\
                & \hspace{2em}\left.
                - 
    273 f_S^3 + f_S^5) + 
 9 f_B^8 (-1 + f_S)^3 (87 + 96 f_S + 124 f_S^2 - 
    250 f_S^3 - 27 f_S^4 + 18 f_S^5) + 
 3 f_B^7 (-1 + f_S)^2 (921 \right.
                \nonumber\\
                & \hspace{2em}\left. 
                - 5782 f_S + 7581 f_S^2 - 
    7268 f_S^3 + 6123 f_S^4 - 1446 f_S^5 + 63 f_S^6) + 
 f_B^5 (60060 - 306980 f_S + 685662 f_S^2 - 771836 f_S^3
 \right.
                \nonumber\\
                & \hspace{2em}\left. 
                +  423070 f_S^4 - 95500 f_S^5 + 3178 f_S^6 + 1260 f_S^7 - 
    66 f_S^8) + 
 f_B^6 (25230 - 141257 f_S + 348325 f_S^2 - 501473 f_S^3
                \right.
                \nonumber\\
                & \hspace{2em}\left.
                 + 
    427073 f_S^4 - 194707 f_S^5 + 39335 f_S^6 - 2499 f_S^7 - 
    27 f_S^8) + 
 f_B^4 (59529 - 283361 f_S + 456189 f_S^2 - 253885 f_S^3 
                \right.
                \nonumber\\
                & \hspace{2em}\left.
                - 
    15645 f_S^4 + 54653 f_S^5 - 15265 f_S^6 + 1249 f_S^7 - 
    8 f_S^8) + 
 f_B^2 (11016 + 40185 f_S - 177621 f_S^2 + 197161 f_S^3 - 
    87965 f_S^4 
                \right.
                \nonumber\\
                & \hspace{2em}\left.
                + 17235 f_S^5 - 1303 f_S^6 + 11 f_S^7 + 
    f_S^8) + 
 f_B^3 (26703 - 71064 f_S - 52318 f_S^2 + 230692 f_S^3 - 
    189036 f_S^4 + 61632 f_S^5 
                \right.
                \nonumber\\
                & \hspace{2em}\left.
                - 8202 f_S^6 + 308 f_S^7 + 
    5 f_S^8)\frac{}{}\right)
    \\
    W_{12} & = -\frac{4 P \kappa_5 f_S^2}{3r^2(1+3f_B)(2+f_B(1-f_S))^2} 
               \left( 9-f_S+5f_B(1-f_S)-6f_B^2(1-f_S) \right)
    \\
    V_{12} & = \frac{-2P\kappa_5f_S}{3r^3(1+3f_B)^2(2+f_B(1-f_S))^3(3-f_S+f_B(1-3f_S))^2}\cdot
               \nonumber\\
               &\hspace{1em}\cdot
               \left(\frac{}{} 27 f_B^7 (-1 + f_S)^3 (1 + f_S) - 
 2 (-3 + f_S)^2 (-54 + 180 f_S - 23 f_S^2 + f_S^3) + 
 9 f_B^6 (-1 + f_S)^2 (-19 - 15 f_S - 21 f_S^2 
                \right.
                \nonumber\\
                & \hspace{2em}\left.
                + 15 f_S^3) +
  f_B^3 (9709 - 29512 f_S + 37424 f_S^2 - 23938 f_S^3 + 
    6915 f_S^4 - 598 f_S^5) + 
 f_B^2 (11733 - 33209 f_S + 38666 f_S^2
                \right.
                \nonumber\\
                & \hspace{2em}\left.
                - 19922 f_S^3 + 
    3857 f_S^4 - 229 f_S^5) + 
 f_B (5868 - 17997 f_S + 18076 f_S^2 - 6238 f_S^3 + 
    840 f_S^4 - 37 f_S^5) 
                \right.
                \nonumber\\
                & \hspace{2em}\left.
                + 
 3 f_B^5 (38 - 587 f_S + 700 f_S^2 + 146 f_S^3 - 
    306 f_S^4 + 9 f_S^5) - 
 2 f_B^4 (-1645 + 6227 f_S - 7659 f_S^2 + 4957 f_S^3 - 
    1880 f_S^4 
                \right.
                \nonumber\\
                & \hspace{2em}\left.
                + 288 f_S^5)
                               \frac{}{}\right)
\end{align}
\begin{align}
W_{22} & = \frac{f_S}{r(3-f_S)(1+3f_B)(2+f_B(1-f_S))^2}\cdot
               \nonumber\\
               &\hspace{1em}\cdot
                 \left(
                  -4 (-3 + f_S) f_S + 
 3 f_B^4 (-1 + f_S)^2 (3 - 15 f_S + 2 f_S^2) + 
 4 f_B (3 - 4 f_S + 5 f_S^2) + 
 f_B^3 (39 - 80 f_S + 18 f_S^2 + 24 f_S^3 - f_S^4)                 
                \right.
                \nonumber\\
                & \hspace{2em}\left.
                - 
 f_B^2 (-48 + 69 f_S - 5 f_S^2 + 7 f_S^3 + f_S^4)              
                 \right)
    \\
    V_{22} & = \frac{f_S}{r^2(1+3f_B)^2(3-f_S)(2+f_B(1-f_S))^3(3-f_S+f_B(1-3f_S))^2}\cdot
            \nonumber\\
            &\hspace{1em}\cdot\left(\frac{}{}
             54 f_B^9 (-1 + f_S)^4 (1 + f_S) + 
 4 (-3 + f_S)^3 (-12 + 7 f_S + f_S^2) - 
 9 f_B^8 (-1 + f_S)^3 (41 + 9 f_S + 117 f_S^2 - 105 f_S^3 + 
    18 f_S^4)                  
                \right.
                \nonumber\\
                & \hspace{2em}\left.
                - 
 4 f_B (-2457 + 5616 f_S - 4716 f_S^2 + 1666 f_S^3 - 
    251 f_S^4 + 14 f_S^5) - 
 3 f_B^7 (-1 + f_S)^2 (1 - 339 f_S - 1954 f_S^2 + 
    2718 f_S^3                   
                \right.
                \nonumber\\
                & \hspace{2em}\left.
                - 615 f_S^4 + 45 f_S^5) + 
 f_B^2 (25236 - 67251 f_S + 71006 f_S^2 - 37009 f_S^3 + 
    8512 f_S^4 - 765 f_S^5 + 14 f_S^6 + f_S^7) + 
 f_B^3 (22989                   
                \right.
                \nonumber\\
                & \hspace{2em}\left.
                - 77809 f_S + 94013 f_S^2 - 58169 f_S^3 + 
    19751 f_S^4 - 2275 f_S^5 - 49 f_S^6 + 13 f_S^7) + 
 f_B^4 (-1195 - 15084 f_S + 29469 f_S^2                   
                \right.
                \nonumber\\
                & \hspace{2em}\left.
                - 13224 f_S^3 + 
    1015 f_S^4 + 724 f_S^5 - 617 f_S^6 + 64 f_S^7) + 
 2 f_B^5 (-5978 + 14642 f_S - 14269 f_S^2 + 13497 f_S^3 - 
    11896 f_S^4                   
                \right.
                \nonumber\\
                & \hspace{2em}\left.
                + 4688 f_S^5 - 753 f_S^6 + 69 f_S^7) + 
 f_B^6 (-5146 + 17301 f_S - 12032 f_S^2 - 5489 f_S^3 + 
    4038 f_S^4 + 2123 f_S^5 - 876 f_S^6 + 81 f_S^7)
                                    \frac{}{}\right)
    \\
    W_{21} & = -\frac{4 P \kappa_5 f_B^2f_S^2}{r^2(1+3f_B)(3-f_S)^2(2+f_B(1-f_S))^2}\cdot
            \nonumber\\
            &\hspace{1em}\cdot
               \left(\frac{}{} 10 + 3 f_B^3 (-9 + f_S) (-1 + f_S)^2 + 8 f_S - 2 f_S^2 + 
  4 f_B (-6 - 7 f_S + f_S^2)                    
                + 
  f_B^2 (-31 + 11 f_S + 23 f_S^2 - 3 f_S^3) \frac{}{}\right)
    \\
    V_{21} & = -\frac{2P\kappa_5f_Bf_S}{r^3(1+3f_B)^2(2+f_B(1-f_S))^3(3-f_S)^3(3-f_S+f_B(1-3f_S))^2}\cdot
            \nonumber\\
            &\hspace{1em}\cdot\left(\frac{}{}
             16 (-3 + f_S)^3 f_S (-5 - 4 f_S + f_S^2) + 
 54 f_B^9 (-1 + f_S)^4 (-3 - 2 f_S + f_S^2) - 
 4 f_B (-3 + f_S)^2 (60 - 11 f_S + 187 f_S^2                    
                \right.
                \nonumber\\
                & \hspace{2em}\left.
                + 117 f_S^3 - 
    43 f_S^4 + 2 f_S^5) - 
 9 f_B^8 (-1 + f_S)^3 (-129 + 142 f_S - 976 f_S^2 + 
    1430 f_S^3 - 423 f_S^4 + 36 f_S^5) + 
 3 f_B^7 (-1 + f_S)^2                    
                \right.
                \nonumber\\
                & \hspace{2em}\left.
                (-189 + 2248 f_S - 18617 f_S^2 + 
    25716 f_S^3 - 7091 f_S^4 + 420 f_S^5 + 9 f_S^6) + 
 2 f_B^2 (-4968 + 174 f_S + 7727 f_S^2 - 1370 f_S^3                    
                \right.
                \nonumber\\
                & \hspace{2em}\left.
                + 
    1843 f_S^4 - 2014 f_S^5 + 581 f_S^6 - 54 f_S^7 + 
    f_S^8) + 
 4 f_B^3 (-867 - 5167 f_S + 31793 f_S^2 - 33910 f_S^3 + 
    14014 f_S^4 - 3911 f_S^5                     
                \right.
                \nonumber\\
                & \hspace{2em}\left.
                + 1005 f_S^6 - 148 f_S^7 + 
    7 f_S^8) + 
 f_B^4 (32142 - 102941 f_S + 246597 f_S^2 - 377773 f_S^3 + 
    249201 f_S^4 - 79311 f_S^5 + 14551 f_S^6                     
                \right.
                \nonumber\\
                & \hspace{2em}\left.
                - 1767 f_S^7 + 
    101 f_S^8) + 
 f_B^5 (38937 - 106226 f_S + 152024 f_S^2 - 270874 f_S^3 + 
    320770 f_S^4 - 158742 f_S^5 + 35856 f_S^6 - 3870 f_S^7        
                \right.
                \nonumber\\
                & \hspace{2em}\left.
                + 
    189 f_S^8) + 
 f_B^6 (13767 - 29608 f_S - 53102 f_S^2 + 142280 f_S^3 - 
    45248 f_S^4 - 53128 f_S^5 + 29758 f_S^6 - 5016 f_S^7 + 
    297 f_S^8)
                                                \frac{}{}\right)
\end{align}

\end{widetext}

\clearpage
\bibliographystyle{apsrev4-1}
\bibliography{nonradialstabilityts.bib}

\end{document}